\documentclass[final,5p,times,twocolumn,nopreprintline]{elsarticle}
\usepackage{amsmath,amssymb,amsfonts,dsfont}
\usepackage{graphicx}
\usepackage{dcolumn}
\usepackage[hyperfootnotes=false]{hyperref}
\usepackage{slashed}
\usepackage{balance}
\usepackage{multirow}
\usepackage{multicol}
\usepackage{booktabs,tabulary}

\newcommand{\Fpi}{F_\pi}
\newcommand{\mpi}{M_{\pi}}
\newcommand{\mpii}{M_{\pi^0}}
\newcommand{\ga}{g_A}
\newcommand{\Order}{\mathcal{O}}

\newcommand{\spiN}{\sigma_{\pi N}}
\newcommand{\bspiN}{\bar\sigma_{\pi N}}
\newcommand{\tspiN}{\tilde\sigma_{\pi N}}

\newcommand{\MeV}{\,\text{MeV}}

\newcommand{\eps}{\epsilon}
\newcommand{\beq}{\begin{equation}}
\newcommand{\eeq}{\end{equation}}

\allowdisplaybreaks[1]

\begin{document}
%%%%%%%%%%%%%%%%%%%%%%%%%%%%%%%%%%%%%%%%%%%%%%%%%%%%%%%%%%%%%%%%%%%%%%%%%%%%%

\renewcommand{\theequation}{\arabic{equation}}

\begin{frontmatter}

\title{On the role of isospin violation in the pion--nucleon $\sigma$-term}

\author[Bern]{Martin Hoferichter}
\author[Madrid]{Jacobo Ruiz de Elvira}
\author[HISKP]{Bastian Kubis}
\author[HISKP,Julich]{Ulf-G.\ Mei{\ss}ner}

\address[Bern]{Albert Einstein Center for Fundamental Physics, Institute for Theoretical Physics, University of Bern, Sidlerstrasse 5, 3012 Bern, Switzerland}

\address[Madrid]{Universidad Complutense de Madrid, Facultad de Ciencias F\'isicas,
Departamento de F\'isica Te\'orica and IPARCOS, Plaza de las Ciencias 1, 28040 Madrid, Spain}

\address[HISKP]{Helmholtz--Institut f\"ur Strahlen- und Kernphysik (Theorie) and
   Bethe Center for Theoretical Physics, Universit\"at Bonn, 53115 Bonn, Germany}
   \address[Julich]{Institut f\"ur Kernphysik, Institute for Advanced Simulation and
   J\"ulich Center for Hadron Physics,   Forschungszentrum J\"ulich, 52425  J\"ulich, Germany}

\begin{abstract}
In recent years, a persistent tension between phenomenological and lattice QCD determinations of the pion--nucleon $\sigma$-term $\spiN$ has developed. 
In particular, lattice-QCD calculations have matured to the point that isospin-violating effects need to be included. Here, we point out that the standard conventions adopted in both fields are incompatible, with the data-driven extraction based on the charged-pion mass, but lattice-QCD conventions relying on the mass of the neutral pion to define the isospin limit. The corresponding correction amounts to  $\Delta \spiN=3.1(5)\MeV$ when evaluated in chiral perturbation theory with low-energy constants determined from a Roy--Steiner analysis of pion--nucleon scattering as well as $\spiN$ itself. It reduces the tension with lattice QCD, and should be included in the comparison to phenomenological determinations. We also update the extraction from pionic atoms accounting for the latest measurement of the width of pionic hydrogen, $\spiN=59.0(3.5)\MeV$, and provide the corresponding set of scalar couplings of the nucleon.  
\end{abstract}

\end{frontmatter}

\section{Introduction}
\label{sec:intro}

The pion--nucleon ($\pi N$) $\sigma$-term
\beq
\label{sigma_term_formal_definition}
\spiN=\hat m\langle N| \bar u u+\bar d d|N\rangle,\qquad \hat m=\frac{m_u+m_d}{2},
\eeq 
quantifies the isospin-symmetric part of the nucleon mass generated by the masses of up- and down-quarks. It is a fundamental parameter of low-energy QCD whose precise determination has long proven elusive, given that, experimentally, it is only accessible indirectly via the Cheng--Dashen low-energy theorem~\cite{Cheng:1970mx,Brown:1971pn}, by an analytic continuation of the isoscalar $\pi N$ amplitude into the unphysical region. Such determination based on the partial-wave
analyses from Refs.~\cite{Koch:1980ay,Hohler:1984ux} yielded values around $\spiN\approx 45\MeV$~\cite{Gasser:1988jt,Gasser:1990ce,Gasser:1990ap}, while more recent
partial-wave analyses~\cite{Arndt:2006bf,Workman:2012hx} favored higher
values, e.g., $\spiN=64(8)\MeV$~\cite{Pavan:2001wz}. Extractions solely based on chiral perturbation theory (ChPT) have generally produced values in line with the input used to constrain the low-energy constants (LECs)~\cite{Fettes:2000xg,Alarcon:2011zs}, but controlling the analytic continuation at the few-MeV level requires a careful analysis using dispersion relations. Such work has been performed in the framework of Roy--Steiner
equations~\cite{Ditsche:2012fv,Hoferichter:2012wf,Hoferichter:2015dsa,Hoferichter:2015tha,Hoferichter:2015hva,Hoferichter:2016ocj,Hoferichter:2016duk,Siemens:2016jwj,RuizdeElvira:2017stg,Hoferichter:2018zwu},
culminating in $\spiN=59.1(3.5)\MeV$~\cite{Hoferichter:2015hva} when combined with pionic-atom constraints on the $\pi N$ scattering lengths~\cite{Strauch:2010vu,Hennebach:2014lsa,Baru:2010xn,Baru:2011bw}. As part of this Letter, we present an updated analysis that includes the final PSI result on the width of pionic hydrogen ($\pi H$)~\cite{Hirtl:2021zqf}, see Sec.~\ref{sec:update}, leading to marginal changes compared to Ref.~\cite{Hoferichter:2015hva}. This result has also been confirmed by an analysis of low-energy $\pi N$ cross sections~\cite{RuizdeElvira:2017stg} (including relatively recent data on elastic reactions~\cite{Brack:1989sj,Joram:1995gr,Denz:2005jq} and the charge-exchange process~\cite{Frlez:1997qu,Isenhower:1999aj,Jia:2008rt,Mekterovic:2009kw}), $\spiN=58(5)\MeV$, establishing agreement between pionic-atom constraints and $\pi N$ scattering data at the $5\MeV$ level.\footnote{$\spiN$ has also been extracted from in-medium modifications of the isovector scattering length in pionic atoms with large $Z$~\cite{Weise:2000xp,Weise:2001sg,Friedman:2019zhc}, but the nuclear uncertainties in such determinations are substantial.} 

However, while the situation among phenomenological determinations thus looks consistent, most calculations in lattice QCD have produced significantly lower values~\cite{Durr:2011mp,Bali:2012qs,Durr:2015dna,Yang:2015uis,Abdel-Rehim:2016won,Bali:2016lvx,Yamanaka:2018uud,Alexandrou:2019brg,Borsanyi:2020bpd} (with few exceptions~\cite{Alexandrou:2014sha}), including the recent works~\cite{RQCD:2022xux,Agadjanov:2023jha}, see Ref.~\cite{FlavourLatticeAveragingGroupFLAG:2021npn} for an overview of the present landscape. A possible resolution to this puzzle was presented in Refs.~\cite{Gupta:2021ahb,Gupta:2022aba} in terms of excited-state contamination, based on evidence both from ChPT and 
 direct lattice calculations, with the main conclusion that close to the physical pion mass results can vary appreciably depending on the analysis strategy to account for excited-state effects.\footnote{Another uncertainty can arise from SU(3) assumptions, e.g., Ref.~\cite{Lutz:2023xpi} differs substantially from Ref.~\cite{RQCD:2022xux} despite being based on the same lattice data.} 
 
 In this Letter, we address another subtlety that becomes relevant with the advent of precision calculations in lattice QCD near the physical point, related to the definition of the isospin limit. Going back to Ref.~\cite{Meissner:1997ii}, the phenomenological convention has been to define $\spiN$ in terms of the charged pion mass, with corrections applied both to $\pi N$ scattering lengths ~\cite{Gasser:2002am,Hoferichter:2009ez,Hoferichter:2009gn,Hoferichter:2012bz} and the low-energy theorem~\cite{Hoferichter:2015dsa,Hoferichter:2015hva}, to ensure that isospin-violating (IV) effects, which are known to be sizable in the  isoscalar $\pi N$ amplitude, are consistently included. In general, the motivation for this convention derives from the fact that most scattering amplitudes and form factors are best measured for charged particles, in such a way that uncertainties of isospin-limit quantities become minimized. Unfortunately, similar considerations in lattice QCD lead to a conflicting definition in terms of the mass of the neutral pion, denoted in the following by $\bspiN$. The main point of this Letter is to clarify the two definitions, including an estimate of the correction that emerges in the transition, see Sec.~\ref{sec:IV}. In Sec.~\ref{sec:scalar} we update the resulting scalar couplings of the nucleon~\cite{Crivellin:2013ipa}, which determine the matrix elements for scalar operators in several search channels for physics beyond the Standard Model, from direct-detection searches for dark
matter~\cite{Bottino:1999ei,Bottino:2001dj,Ellis:2008hf,Hoferichter:2016nvd,Hoferichter:2017olk,Hoferichter:2018acd},
 to neutrino scattering~\cite{Altmannshofer:2018xyo,Hoferichter:2020osn},  $\mu\to e$ conversion in
nuclei~\cite{Cirigliano:2009bz,Crivellin:2014cta,Cirigliano:2022ekw,Davidson:2022nnl,Hoferichter:2022mna}, and electric dipole
moments~\cite{Engel:2013lsa,deVries:2015gea,deVries:2016jox,Yamanaka:2017mef}. We summarize our findings in Sec.~\ref{sec:conclusions}.

 \begin{figure}
\centering
\includegraphics[width=\linewidth,clip]{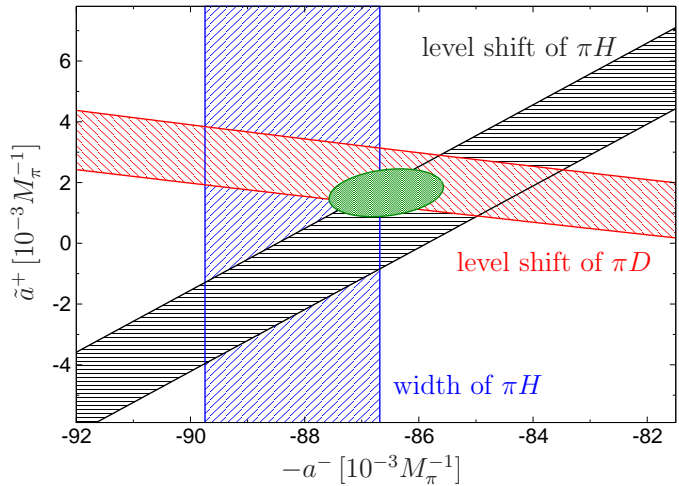}
\caption{Combined constraints on $\tilde a^+$ and $a^-$ from data on the width and level
shift of $\pi H$, as well as the $\pi D$ level shift. The figure is an updated version of the ones from Refs.~\cite{Baru:2010xn,Baru:2011bw,Hoferichter:2015hva} to account for the new value of the $\pi H$ width~\cite{Hirtl:2021zqf}.}
\label{fig:bands}
\end{figure}
 
\section{Update of pionic-atom scattering lengths and $\boldsymbol{\sigma_{\pi N}}$}
\label{sec:update}

In this section we first update the extraction of $\pi N$ scattering lengths from pionic atoms in view of the final result for the width of the ground-state level of pionic hydrogen  from the PSI pionic-atom program~\cite{Hirtl:2021zqf}
\beq
\label{width}
\Gamma_{1s}^{\pi H}=0.856(27)\,\text{eV}.
\eeq
We further use for the level shifts in $\pi H$ and pionic deuterium ($\pi D$)~\cite{Strauch:2010vu,Hennebach:2014lsa}
\beq
\label{level_shifts}
\eps_{1s}^{\pi H}=-7.0858(96)\,\text{eV},\qquad 
\eps_{1s}^{\pi D}=2.356(31)\,\text{eV}.
\eeq
From $\eps_{1s}^{\pi H}$ one obtains, via the improved Deser formula~\cite{Deser:1954vq,Lyubovitskij:2000kk,Gasser:2007zt}, the elastic $\pi^-p\to\pi^- p$ scattering length
\beq
a_{\pi^-p}=85.25(11)\times 10^{-3}\mpi^{-1},
\eeq
with $\mpi$ the {\em charged} pion mass,\footnote{These are the standard units for $\pi N$ scattering lengths, see, e.g., Ref.~\cite{Hohler:1984ux} for a review on $\pi N$ conventions.}
and from $\Gamma_{1s}^{\pi H}$ the charge-exchange $\pi^-p\to\pi^0 n$ analog
\beq
\label{piH_cex}
a_{\pi^-p}^\text{cex}=-124.3(2.0)\times 10^{-3}\mpi^{-1},
\eeq
where both scattering lengths include the contributions from virtual photons.  Using the IV corrections $\Delta a_{\pi^-p}^\text{cex}=0.4(9)\times 10^{-3}\mpi^{-1}$~\cite{Hoferichter:2009ez,Hoferichter:2009gn}, the constraint~\eqref{piH_cex} translates to the isovector scattering length
\beq
\label{aminus_width}
a^-=88.2(1.4)_\text{exp}(0.6)_\text{th}\times 10^{-3}\mpi^{-1}=88.2(1.5)\times 10^{-3}\mpi^{-1}.
\eeq

In contrast, the $\pi D$ level shift does not constrain a single combination of scattering lengths, with many-body corrections that need to be considered~\cite{Weinberg:1992yk,Beane:2002wk,Baru:2004kw,Lensky:2006wd,Baru:2007wf,Liebig:2010ki,Baru:2012iv}, to the effect that the resulting constraint is best analyzed in the context of a global fit, see Fig.~\ref{fig:bands}. 
Following the strategy in Refs.~\cite{Baru:2010xn,Baru:2011bw}, we obtain
\beq
\label{bands}
\tilde a^+=1.7(8)\times 10^{-3} \mpi^{-1},\qquad a^-=86.6(1.0)\times 10^{-3}\mpi^{-1},
\eeq
and a correlation coefficient $\rho_{a^-\tilde a^+}=-0.24$, where the isoscalar scattering length $\tilde a^+$ absorbs the dominant IV corrections.  
The resulting value for $a^-$ is almost identical to the naive average $a^-=86.5(1.0)\times 10^{-3}\mpi^{-1}$ of Eq.~\eqref{aminus_width} and the result if the width of $\pi H$ were dropped altogether
\beq
\label{bands_no_width}
\tilde a^+=1.9(8)\times 10^{-3} \mpi^{-1},\qquad a^-=85.3(1.3)\times 10^{-3}\mpi^{-1},
\eeq
reflecting a slight $1.4\sigma$ 
tension in $a^-$. The isovector channel is intimately related to the $\pi N$ coupling constant $g_c$ via the Goldberger--Miyazawa--Oehme (GMO) sum rule~\cite{Goldberger:1955zza}; for completeness, we provide the updated analysis in~\ref{app:GMO}.  

\begin{table}
\centering
\scalebox{0.747}{
\begin{tabular}{ccrcr}
\toprule
isospin limit & channel & scattering length & channel & scattering length \\\midrule
$a^++a^-$ & $\pi^-p\rightarrow \pi^-p$ & $86.3(1.8)$ & $\pi^+n\rightarrow \pi^+n$ & $85.4(1.8)$ \\
$a^+-a^-$ & $\pi^+p\rightarrow \pi^+p$ & $-88.8(1.8)$ & 	$\pi^-n\rightarrow \pi^-n$  & $-89.7(1.8)$ \\
$-\sqrt{2}\,a^-$ & $\pi^-p\rightarrow \pi^0n$ & $-122.0(1.7)$ &  	$\pi^+n\rightarrow \pi^0 p$ & $-120.1(1.7$\\
$a^+$ & $\pi^0p\rightarrow \pi^0p$ & $2.2(2.6)$ & 		$\pi^0n\rightarrow \pi^0 n$ &  $5.6(2.6)$\\\midrule
$a^++a^-$ & $\pi^-p\rightarrow \pi^-p$ & $85.25(11)$ & $\pi^+n\rightarrow \pi^+n$ & $84.4(7)$ \\
$a^+-a^-$ & $\pi^+p\rightarrow \pi^+p$ & $-87.8(1.6)$ & 	$\pi^-n\rightarrow \pi^-n$  & $-88.6(1.7)$ \\\bottomrule
\end{tabular}}
\caption{$\pi N$ scattering lengths for the physical channels in units of $10^{-3}\mpi^{-1}$, including virtual photons. The upper panel refers to the results based on $\tilde a^+$, $a^-$, and $a^+$, while in the lower panel $a^-$ is eliminated in favor of $a_{\pi^-p}$. Updated from Ref.~\cite{Hoferichter:2015hva}.}
\label{table:physical_channels}
\end{table}

With $\tilde a^+$ and $a^-$ determined from the global fit~\eqref{bands}, the physical scattering lengths follow as given in Table~\ref{table:physical_channels} once combined with the IV corrections. In particular, the lower panel also shows results for the scenario in which $a^-$ is eliminated in favor of $a_{\pi^-p}$. Due to the shift in $a^-$ as implied by the new $\pi H$ width, the central values differ a little more than before, but within uncertainties agreement is still good (even for $a_{\pi^-p}^\text{cex}$ the difference to Eq.~\eqref{piH_cex} is less than $1\sigma$). The isoscalar scattering length itself is updated to
\beq
a^+=7.8(2.6)\times 10^{-3}\mpi^{-1}.
\eeq
For completeness, the results for the channels $\pi^\pm p\to\pi^\pm p$, $\pi^-p\to\pi^0 n$ can also be compared to direct extractions from low-energy cross section data~\cite{RuizdeElvira:2017stg}
\begin{align}
a_{\pi^-p}&=83.3(2.1)\times 10^{-3}\mpi^{-1},\notag\\
a_{\pi^+p}&=-85.7(4.2)\times 10^{-3}\mpi^{-1},\notag\\
a_{\pi^-p}^\text{cex}&=-122.6(4.1)\times 10^{-3}\mpi^{-1},
\end{align}
where we included virtual-photon effects following the treatment of radiative corrections in Ref.~\cite{RuizdeElvira:2017stg}. Within uncertainties, these numbers are fully consistent with the pionic-atom data, albeit with larger uncertainties.

Finally, the scattering length relevant for $\spiN$ is the isoscalar combination, so that the impact of the new $\pi H$ width is limited, but there are still indirect effects due to the global fit. In particular, the master formula in Refs.~\cite{Hoferichter:2015hva,Hoferichter:2015dsa,Hoferichter:2016ocj} is formulated in terms of the virtual-photon-subtracted scattering lengths in isospin basis
\beq
 a^{1/2}_0=169.8(2.0)\times 10^{-3}\mpi^{-1},\qquad  a^{3/2}_0=-86.3(1.8)\times 10^{-3}\mpi^{-1},
\eeq
which are the reference values in
\beq
\label{master_sigma}
\spiN=59.1(3.1)\MeV +\sum_{I_s=1/2,3/2}c_{I_s}\big(a^{I_s}-a^{I_s}_0\big),
\eeq
with coefficients $c_{1/2}=0.242\MeV\times 10^3\mpi$, $c_{3/2}=0.874\MeV\times 10^3\mpi$,
leading to  $\spiN=59.1(3.5)\MeV$. Naively rotating back to $I=\pm$ basis would produce coefficients $c_{+}=1.116\MeV\times 10^3\mpi$, $c_{-}=-0.390\MeV\times 10^3\mpi$, so that the change of $\Delta a^-=0.6\times 10^{-3}\mpi^{-1}$, $\Delta a^+=-0.1\times 10^{-3}\mpi^{-1}$ due to the $\pi H$ update would lower $\spiN$ by $0.3\MeV$.
Following the previous strategy~\cite{Hoferichter:2015hva,Hoferichter:2015dsa} to replace all scattering lengths in terms of $a_{\pi^-p}$ and $\tilde a^+$, we find
\beq
 a^{1/2}=169.9(2.0)\times 10^{-3}\mpi^{-1},\qquad  a^{3/2}=-86.5(1.8)\times 10^{-3}\mpi^{-1},
\eeq
and thereby a shift of $-0.15\MeV$, demonstrating that for $\spiN$ it does not matter how the mild tension in $a^-$ is treated. For the final updated pionic-atom-based value we quote
\beq
\label{final_piN_atomic_atoms}
\spiN=59.0(3.5)\MeV.
\eeq

\section{Isospin violation and $\boldsymbol{\spiN}$}
\label{sec:IV}

To clarify the role of IV in $\spiN$ we start from the chiral expansion of the nucleon mass~\cite{Meissner:1997ii,Muller:1999ww}
\begin{align}
\label{nucleon_mass}
m_N&=m_0-4c_1\mpii^2-\frac{e^2\Fpi^2}{2}(f_1\pm f_2+f_3)\\
&\pm2Bc_5(m_d-m_u)-\frac{\ga^2\big(2\mpi^3+\mpii^3\big)}{32\pi \Fpi^2}+\Order\big(\mpi^4\big),\notag
\end{align}
where the upper/lower sign refers to proton/neutron, and $\mpi$ denotes the mass of the charged pion, i.e.,
\begin{align}
\label{pion_mass}
\mpi^2&=B(m_u+m_d)+2e^2\Fpi^2Z+\Order(m_q^2),\notag\\
 \mpii^2&=B(m_u+m_d)+\Order(m_q^2),
\end{align}
at leading order in the chiral expansion. The scalar couplings $f_q^N$,
\beq
\langle N|m_{q} \bar q q| N\rangle=f_q^N m_N,
\eeq
follow from the Feynman--Hellmann theorem~\cite{Hellmann:1937,Feynman:1939zza}, with the result~\cite{Crivellin:2013ipa}
\begin{align}
\label{scalar_couplings}
 f_u^N & = -\frac{2B}{m_N}
     m_u
  \Big[
     2c_1\pm c_5
    +\frac{3\ga^2(2\mpi+\mpii)}{128\pi\Fpi^2}\Big]+\Order\big(\mpi^4\big), \notag \\
f_d^N & = -\frac{2B}{m_N}     m_d
  \Big[
     2c_1\mp c_5
       +\frac{3\ga^2(2\mpi+\mpii)}{128\pi\Fpi^2}\Big]+\Order\big(\mpi^4\big).   
\end{align} 
Based on these relations, the formal definition in Eq.~\eqref{sigma_term_formal_definition} becomes
\beq
\label{tildesigma}
\hat m\langle N| \bar u u+\bar d d|N\rangle=-4c_1\mpii^2-\frac{3\ga^2\mpii^2(2\mpi+\mpii)}{64\pi\Fpi^2}+\Order(\mpi^4),
\eeq
so that one obtains an expression involving mixed pion charges. The key difference between phenomenological and lattice-QCD conventions is now that $\spiN$ is identified differently in the isospin limit. 

From the phenomenological perspective, the canonical choice~\cite{Meissner:1997ii} amounts to 
\beq
\spiN=-4c_1\mpi^2-\frac{9\ga^2\mpi^3}{64\pi\Fpi^2}+\Order(\mpi^4),
\eeq
because most input data are available for the charged-pion processes: for pionic atoms, the best constraint arises from $a_{\pi^-p}$ via the level shift in $\pi H$, and $\pi D$ is sensitive to the scattering of the charged pion as well. Similarly, most scattering data are available for the elastic processes $\pi^\pm p\to \pi^\pm p$, which motivates the corresponding choice for $\spiN$. To ensure consistency with this definition, the determination of $\spiN$ in Refs.~\cite{Hoferichter:2015hva,Hoferichter:2015dsa} relies on a version of the Cheng--Dashen theorem that accounts for IV corrections, in fact,  a substantial part of the uncertainty in the master formula~\eqref{master_sigma}, $2.2\MeV$, derives from an estimate of LECs parameterizing IV effects. 

In contrast, the prevalent convention in lattice QCD~\cite{FlavourLatticeAveragingGroupFLAG:2021npn} amounts to 
\beq
\bspiN=-4c_1\mpii^2-\frac{9\ga^2\mpii^3}{64\pi\Fpi^2}+\Order(\mpi^4),
\eeq
essentially, to avoid the electromagnetic effects that would otherwise have to be included in the charged pion mass via Eq.~\eqref{pion_mass}. When comparing phenomenological and lattice-QCD determinations, the difference
\beq
\Delta\spiN\equiv\spiN-\bspiN
\eeq
therefore must be taken into account once a level of precision is reached at which IV corrections become relevant. This necessity has already been pointed out in Ref.~\cite{Hoferichter:2016ocj}, but to our knowledge not been considered in subsequent comparisons. 

In practice, quark masses in lattice-QCD simulations are never tuned exactly to the mass of the neutral pion, but the analysis requires a residual chiral  extrapolation or interpolation. Accordingly, to first approximation, $\Delta\spiN$ could be included by evaluating the result at $\mpi$, even though the difference is of electromagnetic origin. Alternatively, an estimate can be obtained from ChPT, by considering the difference of the chiral expansion of the isospin-limit $\spiN$ evaluated at the mass of the charged and neutral pion, respectively. Using the LECs and their correlations from Refs.~\cite{Hoferichter:2015tha,Hoferichter:2015hva}, we obtain 
\beq
\Delta\spiN=\Big\{3.7(1),3.2(1),3.1(1)(4)\Big\}\MeV, 
\eeq
at $\Order\big(p^{2,3,4}\big)$, where the first error propagates the uncertainty in the LECs and the second error at $\Order\big(p^4\big)$ reflects the uncertainty in Eq.~\eqref{final_piN_atomic_atoms}, as required to determine the new LEC $e_1$ at this order in the chiral expansion.  Accordingly, the resulting correction is sizable---we quote
\beq
\label{Deltasigma}
\Delta \spiN=3.1(5)\MeV
\eeq
as our best estimate---and should be included in the comparison of lattice-QCD results with phenomenological determinations.

\section{Scalar couplings}
\label{sec:scalar}

Expressing Eq.~\eqref{scalar_couplings} in terms of $\spiN$, 
\begin{align}
\label{ChPT_res}
 m_N  f_u^N &=\frac{\tspiN}{2}(1-\xi)\pm Bc_5\big(m_d-m_u\big)   \Big(1-\frac{1}{\xi} \Big),\notag\\
 m_N   f_d^N &=\frac{\tspiN}{2}(1+\xi)\pm Bc_5\big(m_d-m_u\big)    \Big(1+\frac{1}{\xi} \Big),
 \end{align}
one encounters a third variant $\tspiN$ as defined by Eq.~\eqref{tildesigma}, 
\begin{align}
 \tspiN&=\spiN+\bigg(4c_1+\frac{21\ga^2\mpi}{128\pi\Fpi^2}\bigg)\Delta_\pi\notag\\
 &=\spiN-3.6(2)\MeV,
\end{align}
where $\Delta_\pi=\mpi^2-\mpii^2$ and~\cite{Giusti:2017dmp,MILC:2018ddw,FlavourLatticeAveragingGroupFLAG:2021npn}
\beq
\xi=\frac{m_d-m_u}{m_d+m_u}=0.365(22).
\eeq
Using, in addition, $B c_5(m_d-m_u)=-0.51(8)\MeV$ as derived from the electromagnetic part of the proton--neutron mass difference $(m_p-m_n)^\text{em}=0.76(30)\MeV$~\cite{Gasser:1974wd},\footnote{The uncertainty is sufficiently large to be consistent with lattice QCD~\cite{Borsanyi:2014jba,Brantley:2016our,CSSM:2019jmq} and more recent evaluations using the Cottingham formula~\cite{Gasser:2015dwa,Gasser:2020mzy,Gasser:2020hzn}.} we find
\begin{align}
\label{scalar_final}
 f^p_u&=19.7(1.4)\times 10^{-3},& f^p_d&=38.3(2.7)\times 10^{-3},\notag\\
 f^n_u&=17.8(1.3)\times 10^{-3},& f^n_d&=42.3(2.6)\times 10^{-3}.
\end{align}
Since the distinction between $\spiN$ and $\tspiN$ was not yet included in Ref.~\cite{Crivellin:2013ipa}, the resulting scalar couplings in Eq.~\eqref{scalar_final} are reduced compared to Ref.~\cite{Hoferichter:2015dsa} (albeit consistent within uncertainties). 

\section{Conclusions}
\label{sec:conclusions}

In this Letter, we addressed a subtlety in the comparison of the pion--nucleon $\sigma$-term as determined phenomenologically from $\pi N$ data and in lattice QCD. While both approaches quote results in the isospin limit, its definition is inconsistent as it proceeds by the mass of charged and neutral pion, respectively. We argued that at the level of maturity reached in recent lattice-QCD calculations, such corrections can no longer be neglected. Estimating the extrapolation from charged to neutral pion mass in chiral perturbation theory, we find a sizable effect, see Eq.~\eqref{Deltasigma}, which reduces the tension between lattice QCD and phenomenology and should be included in future comparisons. We also provided updated results for $\pi N$ scattering lengths and $\spiN$ when including the latest PSI measurement of the width of pionic hydrogen, as well as the scalar couplings of the nucleon, all of which further emphasize the importance of a careful treatment of isospin-violating corrections.

\section*{Acknowledgements}

We thank Harvey Meyer for comments on the manuscript. 
Financial support by the SNSF (Project No.\  PCEFP2\_181117), the DFG through the funds provided to the Sino--German Collaborative
Research Center TRR110 ``Symmetries and the Emergence of Structure in QCD''
(DFG Project-ID 196253076 -- TRR 110), the Ram\'on y Cajal program (RYC2019-027605-I) of the Spanish MINECO,
 the European Research Council
(ERC) under the European Union's Horizon 2020 research and innovation program (ERC AdG EXOTIC, grant agreement
No.\ 101018170), and the  Volkswagen Stiftung (Grant No.\ 93562) is gratefully acknowledged. 

\appendix

\section{GMO sum rule}
\label{app:GMO}

To determine the $\pi N$ coupling $g_c$ from the scattering lengths via the GMO sum rule, one needs input for the combination $a_{\pi^-p}-a_{\pi^+p}$ and remove the effects of virtual photons still contained in the scattering lengths. The strategy followed in Refs.~\cite{Baru:2010xn,Baru:2011bw,Hoferichter:2015hva} amounts to expressing this combination in terms of $a_{\pi^-p}$, $\tilde a^+$, as well as the correction from $a_{\pi^-n}-a_{\pi^+p}$. This strategy leads to $g_c^2/(4\pi)=13.67(12)(15)=13.67(20)$, where the first and second error refer to scattering lengths and dispersive integral, respectively. Since 
most of the scattering-length input then comes from $\eps_{1s}^{\pi H}$ and $\eps_{1s}^{\pi D}$, it is natural to ask how $g_c^2$ would change if one relied primarily on $a_{\pi^-p}^\text{cex}$ instead. 

To this end, we write 
\beq
a_{\pi^-p}-a_{\pi^+p}=-\sqrt{2}a_{\pi^-p}^\text{cex}-\big[a_{\pi^+p}-a_{\pi^-p}-\sqrt{2} a_{\pi^-p}^\text{cex}\big],
\eeq
where the correction term in brackets gives precisely (half) the numerator of the triangle relation
\beq
R=2\frac{a_{\pi^+p}-a_{\pi^-p}-\sqrt{2} a_{\pi^-p}^\text{cex}}{a_{\pi^+p}-a_{\pi^-p}+\sqrt{2} a_{\pi^-p}^\text{cex}}.
\eeq
If we take $a_{\pi^-p}-a_{\pi^+p}$ as determined from the combination of $\pi H$ and $\pi D$ level shifts alone, see Eq.~\eqref{bands_no_width}, $a_{\pi^-p}-a_{\pi^+p}=172.5(1.7)\times 10^{-3}\mpi^{-1}$, we obtain, apart from the correction for IV in $a_{\pi^-n}-a_{\pi^+p}$, a largely experimental determination
\beq
R=-1.9(1.9)\%,
\eeq
in $1.5\sigma$ tension with the ChPT prediction $R=1.5(1.1)\%$~\cite{Hoferichter:2009ez}.

Turning this around and using the chiral prediction for the IV corrections as input, we obtain for the virtual-photon-subtracted scattering lengths
\begin{align}
a_{\pi^-p}^{\slashed{\gamma}}-a_{\pi^+p}^{\slashed{\gamma}}&=-\sqrt{2}a_{\pi^-p}^\text{cex}+0.7(1.0)\times 10^{-3}\mpi^{-1}\notag\\
&=176.5(2.9)\times 10^{-3}\mpi^{-1},
\end{align}
to be compared with $a_{\pi^-p}^{\slashed{\gamma}}-a_{\pi^+p}^{\slashed{\gamma}}=170.9(2.4)\times 10^{-3}\mpi^{-1}$ from the level-shift-focused strategy discussed above, i.e., $\tilde a^+$ from Eq.~\eqref{bands} and $a_{\pi^-p}$ from $\eps_{1s}^{\pi H}$. This difference, as well as the corresponding coupling constant $g_c^2/(4\pi)=13.96(15)(15)=13.96(22)$, again reflects the same $1.5\sigma$ tension as in the triangle relation.

\bibliographystyle{apsrev4-1_mod}
\balance
\biboptions{sort&compress}
\bibliography{ref}

\end{document}